\documentclass[conference]{IEEEtran}
\IEEEoverridecommandlockouts
\makeatletter
\newif\if@restonecol
\makeatother

\usepackage[linesnumbered,ruled,vlined]{algorithm2e}
\usepackage{algpseudocode}
\usepackage{amsmath}
  
\usepackage{algpseudocode}  
\usepackage{amsmath}
\usepackage{amsmath,amssymb,amsfonts}
\usepackage{supertabular}
\usepackage{textcomp}
\usepackage{xcolor}
\usepackage[hyperref=true,backend=biber,sorting=none,backref=true]{biblatex}
\usepackage{booktabs}
\usepackage{graphicx}
\usepackage{subfigure}
\usepackage{multirow}
\usepackage{multicol}
\usepackage{makecell}
\usepackage{graphicx}
\usepackage{threeparttable}
\usepackage{stfloats}

\def\BibTeX{{\rm B\kern-.05em{\sc i\kern-.025em b}\kern-.08em
    T\kern-.1667em\lower.7ex\hbox{E}\kern-.125emX}}
\begin{document}

\title{Heart-Darts: Classification of Heartbeats Using Differentiable Architecture Search\\
}

\author{\IEEEauthorblockN{Jindi Lv}
\IEEEauthorblockA{
\textit{Sichuan University}\\
China, Chengdu \\
lvjindi@stu.scu.edu.cn}
\and
\IEEEauthorblockN{Qing Ye}
\IEEEauthorblockA{
\textit{Sichuan University}\\
China, Chengdu \\
fuyeking@stu.scu.edu.cn}
\and
\IEEEauthorblockN{Yanan Sun}
\IEEEauthorblockA{
\textit{Sichuan University}\\
China, Chengdu \\
ysun@scu.edu.cn}
\and
\IEEEauthorblockN{Juan Zhao}
\IEEEauthorblockA{
\textit{Sichuan University}\\
China, Chengdu \\
zhaojuan2@stu.scu.edu.cn}
\and
\IEEEauthorblockN{Jiancheng Lv}
\IEEEauthorblockA{
\textit{Sichuan University}\\
China, Chengdu \\
lvjiancheng@scu.edu.cn}
}

\maketitle

\begin{abstract}
Arrhythmia is a cardiovascular disease that manifests irregular heartbeats. In arrhythmia detection, the electrocardiogram (ECG) signal is an  important diagnostic technique.
However, manually evaluating ECG signals is a complicated and time-consuming task. With the application of convolutional neural networks (CNNs), the evaluation process has been accelerated and the performance is improved. 
It is noteworthy that the performance of CNNs heavily depends on their architecture design, which is a complex process grounded on expert experience and trial-and-error.
In this paper, we propose a novel approach, Heart-Darts, to efficiently classify the ECG signals by automatically designing the CNN model with the differentiable architecture search (i.e., Darts, a cell-based neural architecture search method).
Specifically, we initially search a cell architecture by Darts and then customize a novel CNN model for ECG classification based on the obtained cells.
To investigate the efficiency of the proposed method, we evaluate the constructed model on the MIT-BIH arrhythmia database. 
Additionally, the extensibility of the proposed CNN model is validated on two other new databases. Extensive experimental results demonstrate that the proposed method outperforms several state-of-the-art CNN models in ECG classification in terms of both performance and generalization capability.
\end{abstract}


\section{Introduction}
Cardiovascular diseases (CVDs) are the dominant causes of death globally during the past 20 years~\cite{mendis2011global}. 
According to the report from the American Heart Association,
CVDs are responsible for 25$\%$ of annual deaths in the United States.
Arrhythmias are abnormal rhythms in heartbeats belonging to the CVDs.
The most common technique to diagnose arrhythmias is to analyze electrocardiogram (ECG) signals, which consist of three main components: P wave, QRS complex, and T wave. It supports the detection of heart disease by measuring the rhythm and electrical activity of the heart. In ECG analysis, heartbeat classification plays a critical role in determining the effectiveness of arrhythmias detection~\cite{kass2005basis}. As manually classifying heartbeats of long-term ECG recordings is a time-consuming task, the application of computer-aided diagnosis (CAD) algorithms provide essential assistance for diagnosis.

Many CAD algorithms have been applied to assist the classification of heartbeats, which require signal preprocessing, waveform detection, and extraction of hand-crafted features~\cite{sahoo2017multiresolution,li2016ecg, elhaj2016arrhythmia,li2016arrhythmia}.
For instance, Li \textit{et al}.~\cite{li2016ecg} utilized the wavelet packet entropy and random forests for ECG classification. Fatin  \textit{et al.}~\cite{elhaj2016arrhythmia} investigated the representation ability of linear and nonlinear features for ECG signals and proposed a method to combine these features together for better ECG classification.
Li \textit{et al}.~\cite{li2016arrhythmia} employed a multi-domain feature extraction approach and constructed a support vector machine (SVM) classifier to classify the heartbeats. Generally, these hand-designed approaches cannot guarantee reliable robustness and easily tend to over-adapt for specific tasks~\cite{wang2020deep}. In addition, CAD systems typically produce a large number of false positives, resulting in longer evaluation time and more biopsies~\cite{lee2017deep}. 
To address this challenge, deep neural networks (DNNs) were employed in ECG analysis.

The DNN differs from traditional CAD algorithms in that it integrates feature extraction and classification together rather
than using hand-crafted features.
The two dominant DNN algorithms in ECG analysis are long short-term memory (LSTM) and convolutional neural networks (CNNs), which have been widely studied by researchers. 
For instance, Yildirim~\textit{et al.}~\cite{yildirim2018novel} adopted the LSTM algorithm with a new wavelet sequence layer for the classification of ECG signals. Kiranyaz \textit{et al.}~\cite{kiranyaz2015real} proposed an adaptive implementation of the 1-D CNN method to distinguish different heartbeat types. Acharya \textit{et al.}~\cite{acharya2017deep,acharya2017automated} designed two CNN models for morphological analysis of ECG signals.

With classical DNNs such as AlexNet~\cite{krizhevsky2017imagenet}, GooleNet~\cite{szegedy2015going}, and ResNet~\cite{he2016deep}, being extensively applied in computer vision (CV) for their remarkable network architecture, the interest is shifting from feature design to architecture design. However, promising architectures in CV do not necessarily have the same good performance for signal processing. Specifically, the CV task deals with images with multidimensional spatial information, whereas the ECG analysis task deals with temporal signals. Applying high-performance networks from CV directly to the ECG analysis task is a process of model migration, which may result in domain bias. Additionally, the process of network architecture design for a specific task relies on trial-and-error experiments and human experience, which is a time-consuming task. 
Fortunately, neural architecture search (NAS), a technique automatically to discover the efficient architectures of the DNNs for a specific task, is proposed and gaining ground to effectively address the aforementioned problems. NAS aims to design efficient and promising DNN architectures based on well-designed search methods such as reinforce learning method (RL), evolutionary algorithm (EA). As the EA-based NAS~\cite{Ye} and RL-based NAS have a high demand for calculation despite the excellent performance, the differentiable architecture search (Darts)~\cite{Liu2018DARTS} was emerged, which is more efficient and less demanding on resources. Instead of searching in a discrete architecture candidate set, Darts searches in a continuous search space and is optimized by gradient descent algorithm.

Darts has been employed with great success in various fields as a powerful and resource-friendly automation tool for automated network architecture design. Inspired by it, we propose a novel approach, name Heart-Darts, to explore high-performance CNN model for ECG classification based on Darts. 
To the best of our knowledge, it is the first work presenting the NAS approach to the ECG classification. 
Specifically, we have also customized a CNN model with a novel architecture for ECG classification. 
To ensure the model be flexibly applied in the ECG analysis, we have validated the generalization capability of the proposed model on two new databases. Extensive experiments have also been conducted to demonstrate the efficiency of the proposed method. The contributions of the proposed Heart-Darts method are shown below:

\begin{itemize}

    \item To explore the impact of CNN network architecture on ECG classification, we introduce Darts, a cell-based neural architecture search technique, as a powerful tool for ECG analysis for the first time.

    \item To further enhance the performance of ECG classification, we design a novel model architecture, i.e., a special combination method of cells in which the cellular architecture is obtained by Darts.
   
   \item To investigate the generalization capability of the proposed model, we migrate the raised model to fresh ECG databases and obtain promising results.
\end{itemize}

The remainder of this paper is organized as follows. The related work and background are discussed in Section~\ref{sec:related work}.
Section~\ref{sec:methodology} elaborates the proposed Heart-Darts method.
Section~\ref{sec:experiments} describes the experimental details, including database information, data preprocessing and parameter settings. The experimental results are presented and analyzed in Section~\ref{sec:result}.
Finally, the conclusions and future work are drawn in Section~\ref{sec:conclusion}. 

\section{Related Work and Background}
\label{sec:related work}
 
In the past decade, a large number of algorithms have been designed for classifying ECG. These algorithms are broadly grouped into two different categories.
The first type of algorithm performs manual feature extraction and classification separately. Conversely, the second type of algorithms typically exploits the end-to-end nature of DNNs~\cite{lei-etal-2018-sequicity} to automate feature extraction followed by the classification. 

The statistical method is one of the representative algorithms in the first category, which relies on manual feature extraction and is prevalent in ECG analysis. For example, Martis \textit{et al.}~\cite{martis2013cardiac} employed principal component analysis (PCA) for dimensionality reduction. Additionally, they utilized a four-layer feed-forward neural network and least square-support vector machine to classify ECG. The method yielded an accurate classification of 93.48$\%$.
Elhaj \textit{et al.}~\cite{elhaj2016arrhythmia} utilized PCA with the discrete wavelet transform coefficients followed by SVM classifier and neural network for classification with an accuracy of 98.91$\%$. Sahoo \textit{et al.}~\cite{sahoo2017multiresolution} applied wavelet transform to detect QRS complexes of ECG signals and extracted information by neural network and SVM classifier, where the diagnostic accuracy was 99.85$\%$. 
Although the statistical methods have made considerable progress in ECG classification, their scalability is poor.

With the advent of DNNs, which is an end-to-end techinique that automatically combines feature extraction and classification together, a large number of works have been built on it for ECG analysis. Particularly, Yildirim \textit{et al.}~\cite{yildirim2018novel} designed two models, i.e., DBLSTM-WS and DULSTM-WS, to classify ECG signals using the LSTM algorithm and achieved recognition rates of 99.39$\%$ and 99.25$\%$, respectively. In their another work, Yildirim \textit{et al.}~\cite{yildirim2019new} founded a CAE-LSTM model that significantly improved the time cost of the LSTM network and obtained an accuracy of 99.23$\%$. 
The application of DNNs in ECG classification has achieved promising success, but the design of specific network architecture has become another major challenge. 

Considering the rising demand for network architecture engineering, NAS came into being. The RL-based NAS~\cite{zoph2016neural} is considered a pioneering work in NAS. They obtained the network architecture using the reinforce learning method and achieved state-of-the-art performance on image classification. Subsequently, the EA-based NAS~\cite{real2019regularized} utilized the evolution algorithm to achieve similar results. However, they both have a high demand for calculation resources. Therefore, numerous works have arisen on reducing the computation resource and accelerating the network architecture search.
The most notable work is the Darts approach proposed by Liu \textit{et al.}~\cite{Liu2018DARTS}, which introduced a novel idea of searching architecture in a continuous search space and achieved state-of-the-art performance. Given the strengths of Darts, the idea of applying Darts to address the aforementioned challenges of ECG classification is motivated.

\section{Methodology}
\label{sec:methodology}
Task-specific network architecture design is a challenging task, which has inspired us to consider using NAS to automate this process for ECG classification. In view of the limitation of computation resources, Darts, an approach of NAS with lower computational resource requirements, becomes a better choice. 
Additionally, the depth of the DNNs is an important factor in determining the performance of the network. In general, the learning ability of DNNs will be improved as the number of network layers increases.
But in fact, simply stacking the number of network layers may cause network performance degradation, since the deep network model will encounter gradient vanishing/explosion problem when the number of network layers is deep enough. To address this issue, ResNet~\cite{he2016deep} proposes an efficient strategy of skipping multiple network layers with ``shortcut connections''. 
Based on the above inspiration, in this paper, we propose a novel method named Heart-Darts for effective classification of ECG signals. 
\begin{figure}[bp]
    \centering
    \includegraphics[width=3.0in]{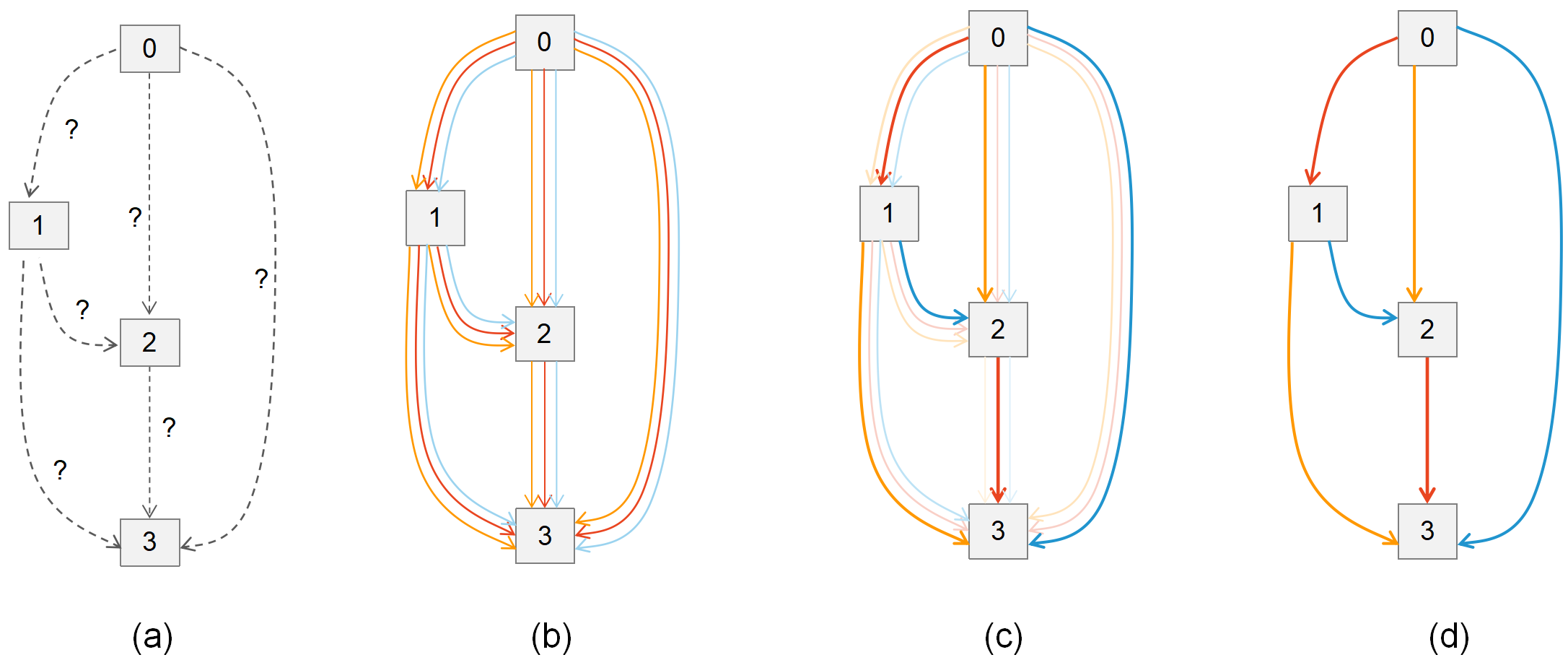}
    \caption{ An overview of Darts: (a) Operations on each edge are initially unknown. (b) A mixture of candidate operations are placed on each edge. (c) Optimization with gradient descent. (d) Operation with max probabilities is selected for each edge to form the final architecture.~\cite{Liu2018DARTS}.}
    \label{fig:Darts}
\end{figure}

\subsection{The proposed Heart-Darts Method}

The Heart-Darts method aims to customize a high-performance model for ECG classification. It mainly consists of two parts: (1) Darts-based cell architecture search (2) cells-based novel model construction. The specific algorithm of Heart-Darts is presented in Algorithm~\ref{alg:Heart-Darts}.

The cell architecture is dominant because of the final model formed by stacking cells. To obtain the optimal cell architecture for ECG classification, we employ the Darts as a powerful tool.   
It introduces a novel idea of searching for the cell architecture in a continuous search space, unlike traditional NAS whose search space is discrete such as EA-based NAS~\cite{real2019regularized} or RL-based NAS~\cite{zoph2016neural}. 

In Darts, a cell is defined as a directed acyclic graph that contains N ordered nodes, where the cells are of two types: normal cell and reduction cell. Specifically, each node $x^{(i)}$ is a latent representation of a feature map computed by Equation (\ref{x(j)}) and each directed edge $(i, j)$ stands for some operations $o^{(i,j)}$ on $x^{(i)}$~\cite{Liu2018DARTS}. Furthermore, the operation candidate set is denoted as $O$, where each operation represents a function $o(\cdot)$ with $x^{(i)}$ as input. To make search space continuous, Darts defines $\alpha=\{\alpha^{(i,j)}\}$ as the encoding of architecture, where $\alpha^{(i,j)}$ is a vector of dimension $|O|$ represents the operation mixing weights for a pair of nodes $(i, j)$. Then $\bar{o}^{(i,j)}$ can be obtained from Equation (\ref{o(i,j)}), which is a weighted summation of all operations on each directed edge $(i, j)$. At the end of the search, $\bar{o}^{(i,j)}$ is replaced with the operation corresponding to the largest $\alpha^{(i,j)}$ value: $o^{(i,j)}=argmax_{o\in O}\alpha_o^{(i,j)}$, and the top-2 strongest operations are selected as inputs for each node.
An overview of Darts is shown in Figure~\ref{fig:Darts}. Since the encoding of cell architecture $\alpha=(\alpha_{normal},\alpha_{reduce})$ is a learnable parameter, the ultimate goal of Darts is to find $\alpha^*$ that minimizes the validation loss by the gradient descent: $\min \limits_{\alpha} L_{val}(\omega,\alpha)$.  It is noteworthy that due to the time consumption issue, we discard the original Darts' bilevel optimization strategy and adopt an alternating update of the weights and $\alpha$. The whole process of cell architecture search is shown in Steps 1 to 6 in Algorithm~\ref{alg:Heart-Darts}.
\begin{equation}
    x^{(j)} = \sum_{i<j}o^{(i,j)}(x^{(i)})
\label{x(j)}
\end{equation}
\begin{equation}
\bar{o}^{(i,j)}(x)=\sum_{o\in O}\frac{exp(\alpha_o^{(i,j)})}{\sum_{o^{'}\in O}exp(\alpha_{o^{'}}^{(i,j)})}o(x)
\label{o(i,j)}
\end{equation}

\begin{figure}[tp]
    \centering
    \includegraphics[width=3.5in]{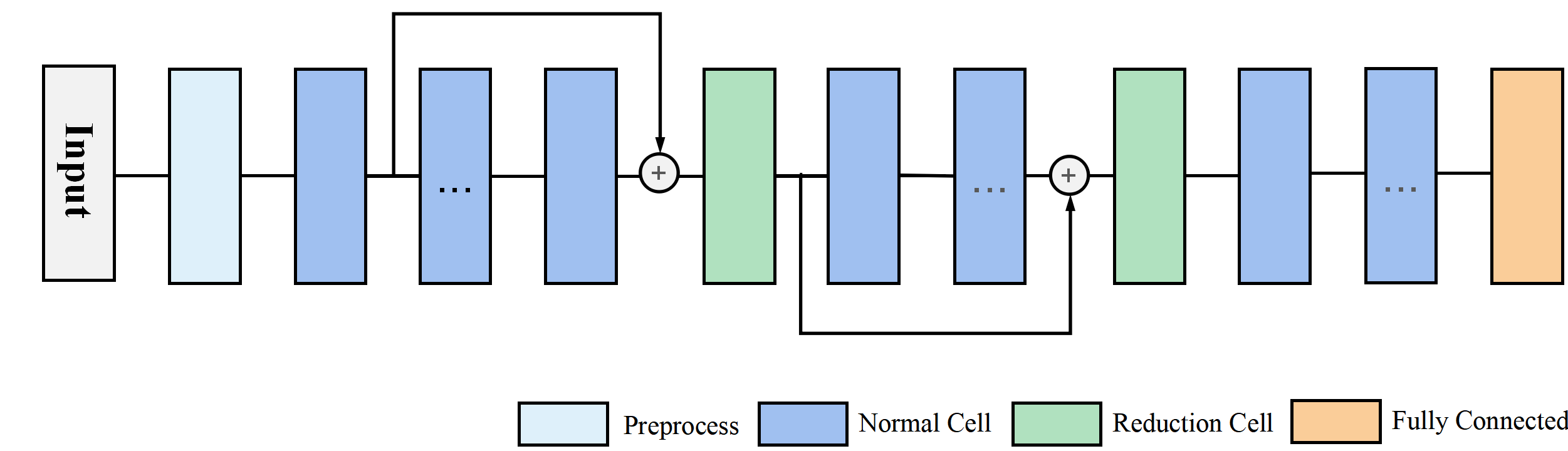}
    \caption{The complete proposed model architecture}
    \label{fig:model_architecture}
\end{figure}

After obtaining the optimal cell architecture, a final model is assembled from the obtained cells, where 1/3 and 2/3 layers are reduction cells follows~\cite{2017Learning,liu2018progressive,real2019regularized}. Although the final model is based on the stack of obtained cells, it does not mean that the more cells the better performance, as gradient vanishing/explosion may occur with the increase of the number of cells. To prevent this phenomenon, we introduce a shortcut strategy of the ResNet~\cite{he2016deep} in the final model architecture. Specifically, we devise a novel model architecture that incorporates a preprocessing layer and two shortcut connections as shown in Fig~\ref{fig:model_architecture}. And considering the dimensional consistency, two skip connections are attached separately before the two reduction cells. In Algorithm~\ref{alg:Heart-Darts}, Steps 8 to 19 show the adding of the shortcut connections to the model constructed in Step 7.

\begin{algorithm}[tp]
  \caption{Heart-Darts algorithm}
  \label{alg:Heart-Darts}
  Define the operation candidate set $O$.\\
  Initialize weights of architecture
  \label{alg:initializaiton}
  $\alpha=(\alpha_{normal},\alpha_{reduce})$.\\
    \While{not converged}
    {
      
      
      Update architecture $\alpha$ by $\nabla_{\alpha}L_{val}(w,\alpha)$\;
      
      Update weights $w$ by $\nabla_{w}L_{train}(w,\alpha)$\; 
    }
  
  Obtain the optimal cell architecture $\alpha^*$
  
  Construct a model without shortcut connections based on the obtained cells.\\
  Initalize input nodes $s_0$, $s_1$.\\
  \While{not converged}{
   
   \For{$i<layers$}{
        compute $s_0$, $s_1 $;\\
        \If{$i==0$ or $i==\frac{layers}{3}$ or $i==\frac{2\times layers}{3}$} {
             $t=s_1;$
        }
        \ElseIf{$i==\frac{layers}{3}-1$ or $i==\frac{2\times layers}{3}-1$}{
        $s_0=ReLu(s_0+t);$\\
        $s_1=ReLu(s_1+t);$
        }
        
   }\
   $output=classifier(s_1)$\\
    Update weights $w$ by descending $\nabla _{w}L_{train}(w)$ 
    }
  \Return the best model
\end{algorithm}

\section{Experiment Design}
\label{sec:experiments}
In this section, we provide details of the chosen databases, data preprocessing and parameter settings used in the experiments for investigating the performance of Heart-Darts.
All experiments are performed on a single 1080Ti GPU with PyTorch implementation.

The experiment is divided into three main stages: cell architecture search, model construction and evaluation, and model generalization validation. The first part focuses on searching for an optimal cell architecture by Darts. Then a novel model is constructed and evaluated based on the obtained cells. In the last part, to validate the generalization capability of the proposed model, we migrate the model on two new database. It is noteworthy that the ordinary cross-entropy function is used in all experiments.

 \begin{figure}[bp]
    \centering
    \includegraphics[width=2.5in]{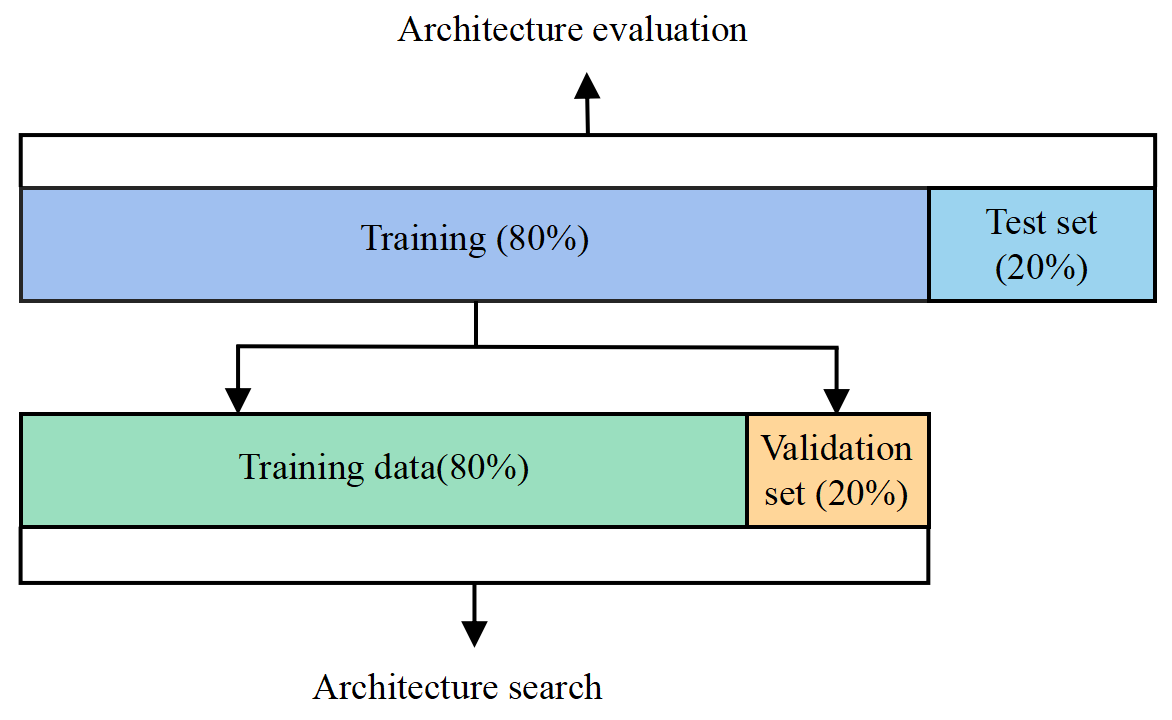}
    \caption{The data distribution details for MIT-BIH arrhythmia database}
    \label{fig:data_distribution}
\end{figure}

\subsection{The Details of Database}\label{sec:database}
We used three databases in this study: MIT-BIH arrhythmia database, St Petersburg Institute of Cardiology Technics(INCART) 12-leads arrhythmia, and QT database. They were obtained from an open-source PhysioNet database\parencite{Goldberger2000PhysioBank}. 
 Among all the data, there are 15 types of heartbeats as shown in Table~\ref{tab:15types}. According to the standards advocated by Advanced Medical Instruments(AAMI)~\cite{AAMI}, the types of heartbeat re-classified into five classes: Normal beats(N), Supraventricular ectopic beats(S), Ventricular ectopic beats(V), Fusion beats(F), Unclassified beats(Q).

The reclassification data are detailed in Table~\ref{tab:classNum}. As can be seen from the table, there is a significant data imbalance problem in each database, especially in the INCART database. The MIT-BIH arrhythmia database, which has a total of 94,013 heartbeats, is involved in the cell architecture search and model evaluation phases. The remaining two databases are availed for validating model generalization. For each database, we randomly select 80\%  as the training set and 20\% as the test set. Regarding the cell architecture search, we need to re-segment the training set of the MIT-BIH arrhythmia database into two parts, with the first 80$\%$ for training and the last 20$\%$ for validation. The data
distribution details of the MIT-BIH arrhythmia database are shown in Figure~\ref{fig:data_distribution}. The details of the three databases are described below.

\begin{figure*}[tp]
\centering
\subfigure{
\includegraphics[width=0.32\textwidth]{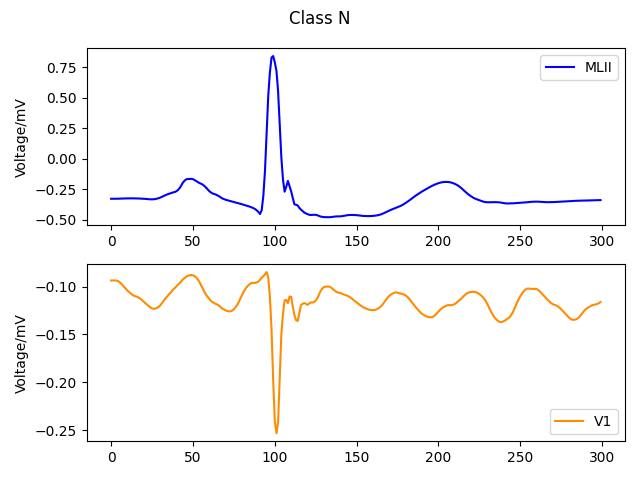}}
\subfigure{
\includegraphics[width=0.32\textwidth]{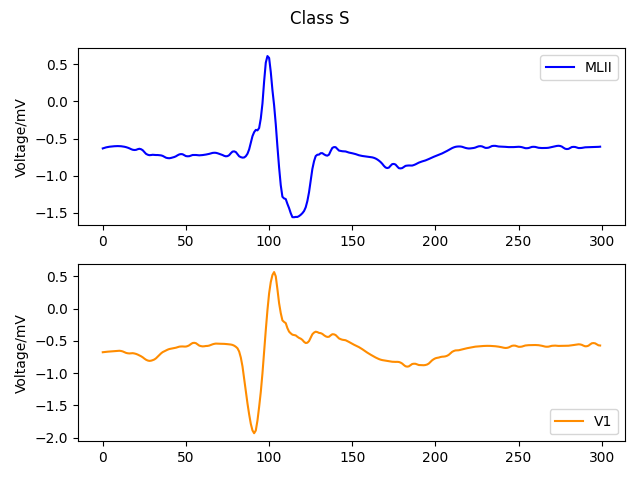}}
\\
\subfigure{
\includegraphics[width=0.32\textwidth]{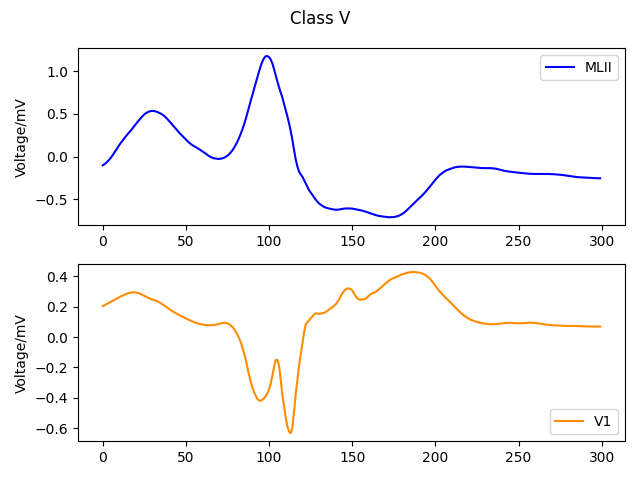}}
\subfigure{
\includegraphics[width=0.32\textwidth]{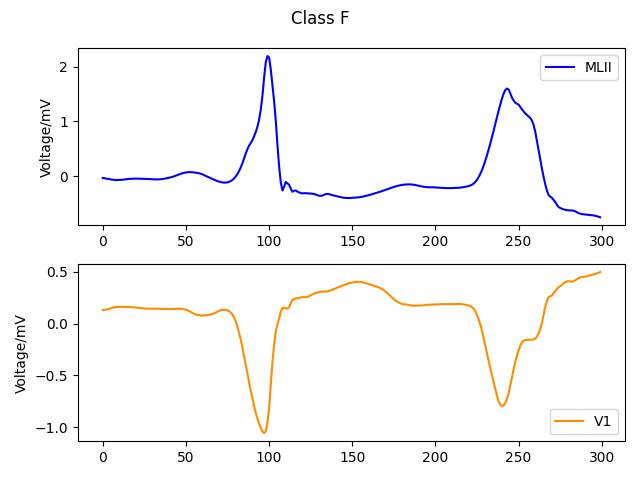}}
\subfigure{
\includegraphics[width=0.32\textwidth]{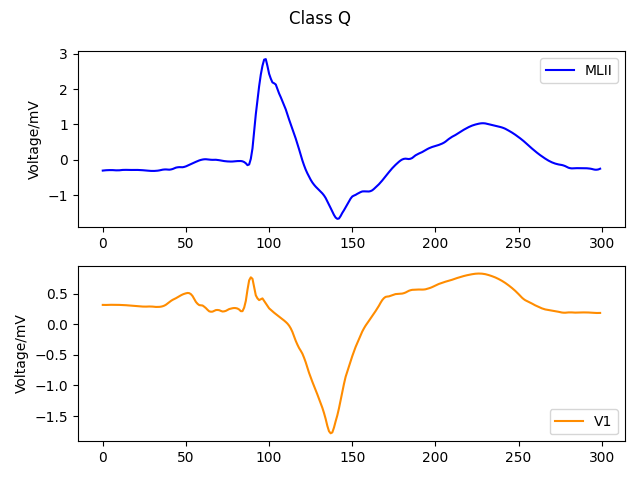}}
\caption{Five types of heartbeats on the MIT-BIH arrhythmia database}
\label{fig:5classes} 
\end{figure*}

\begin{table*}[tp]
\centering
\caption{Fifteen arrhythmia heartbeats in the MIT-BIH database..}
\setlength{\tabcolsep}{9mm}{
\begin{tabular}{llll}
\toprule
Annotation & Heartbeat type &Annotation& Heartbeat type \\
\midrule
N&Normal beat& S& Supraventricular premature beat \\
L &Left bundle branch block beat&j& Nodal (junctional) escape beat \\
R& Right bundle branch block beat& f& Fusion of paced and normal beat\\
A & Atrial premature beat&E&  Ventricular escape beat \\
V& Premature ventricular contraction& J& Nodal (junctional) premature beat\\
a& Aberrated atrial premature beat& e& Atrial escape beat\\
 P& Paced beat&Q& Unclassifiable beat\\
F&Fusion of ventricular and normal beat  \\
\bottomrule
\end{tabular}}
\label{tab:15types}
\end{table*}

\begin{table*}[tp]
    \centering
    \caption{The five classes of division schemes recommended by AAMI.}
    \setlength{\tabcolsep}{3.5mm}{
    \begin{tabular}{p{1.0cm}p{2.8cm} p{4.0cm} p{2.0cm} p{2.0cm}p{2.0cm}}
    \toprule
    AAMI class& MIT-BIH classes annotations&Group description database& MIT-BIH database& INCART database& QT databse    \\
    \midrule
    N& L, N, R, e, j& Normal beats& 79,508& 153,350& 81,022\\ 
    S& A, J, S, a& Supraventricular ectopic beats& 2696& 1957& 1485\\ 
    V &E, V &Ventricular ectopic beats& 7129& 19,973& 1691\\
    F &F &Fusion beats &793& 219 & 251\\
    Q& P, Q, f& Unclassified beats& 3887& - & 2146 \\
    Total& 15& 5& 94,013& 175,499&86,595\\
    \bottomrule
    \end{tabular}}
    \label{tab:classNum}
\end{table*}

\begin{table*}[tp]
\centering
\caption{Distribution of lead types and record numbers in MIT-BIH arrhythmia database.}
\setlength{\tabcolsep}{2.5mm}{
\begin{tabular}{lllllll}

\toprule
Lead A& V5& V5& MLII& MLII& MLII& MLII\\
Lead B& V2& MLII& V2& V4& V5& V1\\
Records& 102& 114&  103& 124& 100 123& 101 105 106 107 108 109 111 112 113 115 116 118 119 121 122 200 201 202 203 205    \\
& 104& & 117& & 123&207 208 209 210 212 213 214 215 217 219 220 221 222 223 228 230 231 232 233 234\\
\bottomrule
\end{tabular}}
\label{tab:MIT-BIH}
\end{table*}

\begin{itemize}
    \item MIT-BIH arrhythmia database contains 48 records obtained from 47 subjects (25 males and 22 females). Each record consists of 30-minutes of dual-channel (noted as lead A and lead B) ambulatory ECG signals with a sampling rate of 360 Hz. In lead A, modified-lead II (MLII) contains 45 records except for 102, 104, 114 records. In lead B, lead V1 involves most of the records. The details of the lead shown in Table~\ref{tab:MIT-BIH}. 
    In this study, two datasets are selected. One is 2-lead dataset consisting of lead V1 and lead MLII with 40 records, and the other is single-lead dataset consisting of lead V1 with the same records as 2-lead dataset.

 \item St Petersburg INCART database consists of 75 annotated recordings obtained from 32 subjects(17 males and 15 females, aged 18-80). Each record contains 30-minutes 12 standard leads ambulatory ECG signals. The sample rate is 257 Hz. In this database, we select lead II and lead V1 to compose a 2-lead dataset like of the MIT-BIH database described above. 
 
 \item QT database contains 100 records of 15-minute dual-lead (noted as lead MLII and lead V5) ECGs with a sampling rate of 250 Hz.
 For comparison, we select lead MLII as a single-lead dataset.

\end{itemize}

\begin{figure}[htb]
     \centering
     \includegraphics[width=2.6in]{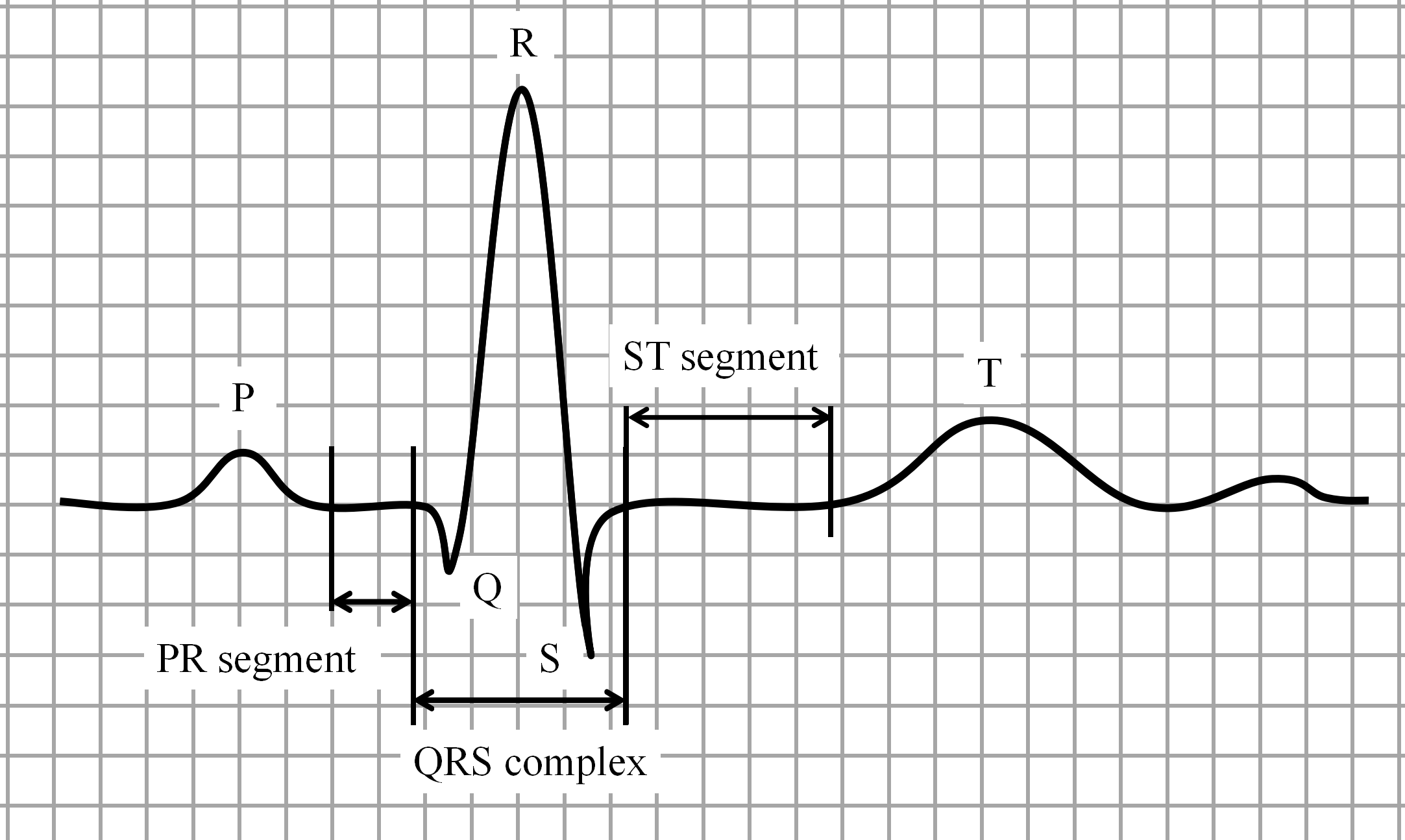}
     \caption{A complete heartbeat observed on an ECG}
     \label{fig:heartbeat}
 \end{figure}

\subsection{The Data Preprocessing}\label{AA}
As the ECG signals are long-term recordings of cardiac activity, it is difficult to directly classify the raw data of the ECG signals. Therefore, a proper data preprocessing should be employed. In this experiment, the data preprocessing consists of two parts: denoising and heartbeat classification.

ECG signals are waveform data associated with time series. The raw ECG signals cover a lot of noise affected by the environment, such as power frequency interference, baseline drift,  high-frequency noises caused by muscle contraction and electrode movement, etc~\cite{Shanshan2017Heartbeat}.
To restore the real characteristics of the ECG signals, denoising is a desirable option. In this study, we use the denoising method based on wavelet transform. 
\begin{table}[bp]
\centering
    \caption{A confusion matrix for single-lead dataset.}
  \scalebox{0.8}{
\begin{threeparttable}
    \begin{tabular}{ccccccccccc}
    \toprule
     &&\multicolumn{5}{c}{Predicted}&+P& Se& Spe&Acc \\
     \cline{3-7}
     &&N&S&V&F&Q & ($\%$)  & ($\%$)&($\%$)&($\%$) \\
    \midrule
     \multirow{5}*{Original}& N & 16007 & 28 & 8 & 1 &  0 & 99.63 & 99.77 & 97.83  & \multirow{5}*{\textbf{99.35}}\\
    &S &  27 &  472 & 5 &	0 &	0 &	94.21 &	93.65 &	99.84 \\
    &V &  15&  1& 1290&	13&	0&	98.47& 97.80& 99.89 \\
    &F	& 14	&0	& 6	&  120	&0	&89.55	&85.71	&99.92\\
    &Q	& 4	&0	&1	&0	&791	&100.00	&99.37	&100.00 \\ 
    \bottomrule
    \end{tabular}
    \begin{tablenotes}
       \footnotesize
       \item[1] Se: Sensitivity, +P: Positive predictivity, Spe: Specificity, Acc: Accuracy.
     \end{tablenotes}
    \label{tab:Heart-Darts-1}
    \end{threeparttable}}
\end{table}

A complete heartbeat consists of P wave, QRS-complex, T wave, PR segment, and ST segment as shown in Figure~\ref{fig:heartbeat}. One record contains multiple heartbeats, which are segmented based on R wave peak position. The standard of segmentation is to make the heartbeat contains the most pathological information. A total of three databases are used in this paper, and the details of the databases are described in Section~\ref{sec:database}. 
For the MIT-BIH arrhythmia database, we choose the first 99 sampling points and the last 200 sampling points of the marked R peak  position. The five types of heartbeats with lead MLII and lead V1 in the MIT-BIH arrhythmia database shown in Figure~\ref{fig:5classes}.
For the INCART database, we choose 99 sampling points before the marked R peak position and 150 sampling points after it. 
For the QT database, we take the first 99 sampling points and the last 120 sampling points of the marked R peak position.


\subsection{Parameter Settings}
In cell architecture search, we work only with the MIT-BIH arrhythmia 2-lead dataset (lead MILL and lead V1). 
According to the previous state-of-the-art network architectures~\cite{zubair2016automated,acharya2017deep, acharya2017automated}, we define the candidate operation set $O$ that includes the following operations: $3\times1$ convolution, $5\times1$ convolution, $9\times1$ convolution, $13\times1$ convolution, $17\times1$ convolution, $27\times1$ convolution, $3\times1$ max pooling,  $5\times1$ max pooling, skip connection and zeros. Operation zero($\cdot$) represents there is no connection between a pair of nodes and the order of the convolution operation is Conv-BN-ReLU. The  preprocessing layer consists of a 1-D convolutional layer (the size of kernels is 5, the strides are 2 and the number of filters is 32), a batch normalization (BN) layer, a ReLU layer, and a max-pooling layer (the pooling size is 3 and the strides are 2).
And the convolutional cell consists of 7 nodes, including two input nodes and one output node. A small network composed of 8 cells is trained by Darts for 100 epochs with a batch size of 100. Additionally, the initial channel of this small network is 16. The rest of the hyperparameters are the same with~\cite{Liu2018DARTS} on CIFAR-10. 

After obtaining the optimal cell architecture in the first stage, a novel CNN model is customized with the 15 cells. To investigate the performance of this model, evaluations are performed on the MIT-BIH arrhythmia single-lead and 2-lead datasets, respectively. We train the model for 100 epochs with a batch size of 256 from scratch and report the performance on the MIT-BIH arrhythmia test dataset. The other hyperparameters are identical to the cell architecture search phase.
Additionally, we train three different CNN models, CNN-9~\cite{acharya2017deep}, CNN-11~\cite{acharya2017automated}, and ResNet-31~\cite{2020Heartbeat}, in the same setups to fairly compare the model performance.  The CNN-9 and CNN-11 are generally regarded as the earliest successful CNN models in ECG classification. The ResNet-31, a block-based network with a novel shortcut connection architecture similar to 
ours, is selected to validate the performance cell architecture.

For the model generalization validation stage, we migrate the proposed model on two new databases: INCART database and QT database. 
The model is trained from scratch with random weights.
All hyperparameters are the same as in the model evaluation phase.
Furthermore, to make a fair comparison, we train the CNN-9 and CNN-11 from scratch with the same settings over the two new databases.

\section{Results Analysis}
\label{sec:result}

\begin{table}[bp]
    \centering
    \caption{A confusion matrix for 2-lead dataset.}
 \scalebox{0.8}{
\begin{threeparttable}
    \begin{tabular}{ccccccccccc}
    \toprule
     &&\multicolumn{5}{c}{Predicted}&+P & Se& Spe&Acc \\
     \cline{3-7}
     &&N&S&V&F&Q& ($\%$)&($\%$)&($\%$)&($\%$) \\
    \midrule
    \multirow{5}*{Original}& N & 16001 &   26 &  7
 &8& 2 & 99.72& 99.73 & 98.37   & 
 \multirow{5}*{\textbf{99.43}}\\
    &S & 27 &   476 & 1 & 0 &	0 &	94.82 &	94.44 &	99.86\\
    &V & 9&	0&	1304&  6&   0&	98.56&	98.86&	99.89 \\ 
    &F	& 7	&0	& 11	& 121	& 1	&89.63	&86.43	&99.92\\
    &Q	&  2&0	&0	&0	&  794	&99.62	&99.75	&99.98 \\ 
    \bottomrule
    \end{tabular}
    \begin{tablenotes}
       \footnotesize
       \item[1] Se: Sensitivity, +P: Positive Predictivity, Spe: Specificity, Acc: Accuracy.
     \end{tablenotes}
    \label{tab:Heart-Darts-2}
    \end{threeparttable}}
\end{table}

Table~\ref{tab:ComparisonCNN} presents the results of the model evaluation on the MIT-BIH arrhythmia test set. 
A comparison is presented between three different CNN models, CNN-9~\cite{acharya2017deep}, CNN-11~\cite{acharya2017automated} and ResNet-31~\cite{2020Heartbeat}, and the proposed Heart-Darts-base model. Additionally, we train an additional Heart-Darts-base model without shortcut connections to verify the proposed shortcut strategy.  
\begin{table}[tp]
    \centering
    \caption{The detailed results of the architecture evaluation.}
   \begin{threeparttable}
    \setlength{\tabcolsep}{2.0mm}{
    \begin{tabular}{llllll}
    \toprule
    Architecture & Lead & Acc &  Se & +P& F1 \\
    &&($\%$)&($\%$)&($\%$)&($\%$)\\
    \midrule
    CNN-9~\cite{acharya2017deep}	&Single-lead	&98.89	& 92.11	&  95.92& 93.98\\
    &	2-Lead&	 99.03&	 93.60&96.43 & 94.99\\
    CNN-11~\cite{acharya2017automated}	&Single-lead &  98.88&	91.38 &96.58 & 93.91\\
    & 2-Lead & 99.10 & 93.15&\textbf{97.29} & 95.18\\
    
    ResNet-31~\cite{2020Heartbeat}	&Single-lead & 99.00 & 91.11 & \textbf{97.05} & 93.98\\
    & 2-Lead & 99.23& 94.31 & 96.51& 95.40\\
    
    Heart-Darts&Single-lead&
   99.29&94.87	& 95.76& 95.31\\
   &2-Lead& 99.39	&95.57	&96.28& 95.92\\
    
   Heart-Darts + shortcut&Single-lead&
   \textbf{99.35}&\textbf{95.26}	& 96.37& \textbf{95.81}\\
   &2-Lead&\textbf{99.43}	&\textbf{95.84}	&96.47& \textbf{96.16}\\
    \bottomrule
    \end{tabular}}
     \begin{tablenotes}
       \footnotesize
       \item[1]Acc: Accuracy, Se: Sensitivity, +P: Positive predictivity, F1: F1-score .
     \end{tablenotes}
    \label{tab:ComparisonCNN}
    \end{threeparttable}
\end{table}

\begin{table}[tp]
    \centering
    \caption{The detailed results of the generalization validation.}
   \begin{threeparttable}
    \setlength{\tabcolsep}{1.9mm}{
    \begin{tabular}{lllllll}
    \toprule
    Architecture& Database & Lead& Acc& Se& +P& F1  \\
    &&($\%$)&($\%$)&($\%$)&($\%$)&($\%$)\\
    \midrule
    CNN-9& QT  &single-lead&99.50&89.59&96.69&93.00\\
    &INCART &2-lead&99.45&77.25&\textbf{95.39}&85.37\\
    
    CNN-11&QT &single-lead&99.36&88.10&97.05&92.36\\
     &INCART &2-lead&99.54&79.42&88.63&83.77\\
     
    Heart-Darts& QT &single-lead&\textbf{99.69}&\textbf{93.76}&\textbf{98.38}&\textbf{96.01}\\
     &INCART &2-lead&\textbf{99.81}&\textbf{90.05}&89.49&\textbf{89.77}\\
    \bottomrule
    \end{tabular}}
    \begin{tablenotes}
       \footnotesize
       \item[1]Acc: Accuracy, Se: Sensitivity, +P: Positive predictivity, F1: F1-score .
     \end{tablenotes}
    \label{tab:generalization}
    \end{threeparttable}
\end{table}

\begin{figure}[bp]
    \centering
    \includegraphics[width=3.0in]{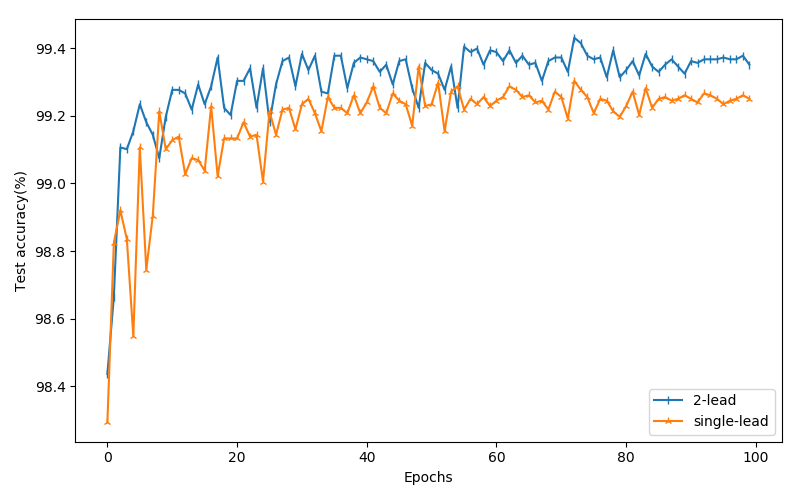}
    \caption{The test accuracy of Heart-Darts on single-lead and 2-lead datasets}
    \label{fig:myplot}
\end{figure}

\begin{figure}[bp]
\centering
\subfigure[The normal cell architecture of 2-lead data]{
\label{fig:2_lead_normal} 
\includegraphics[width=3.0in]{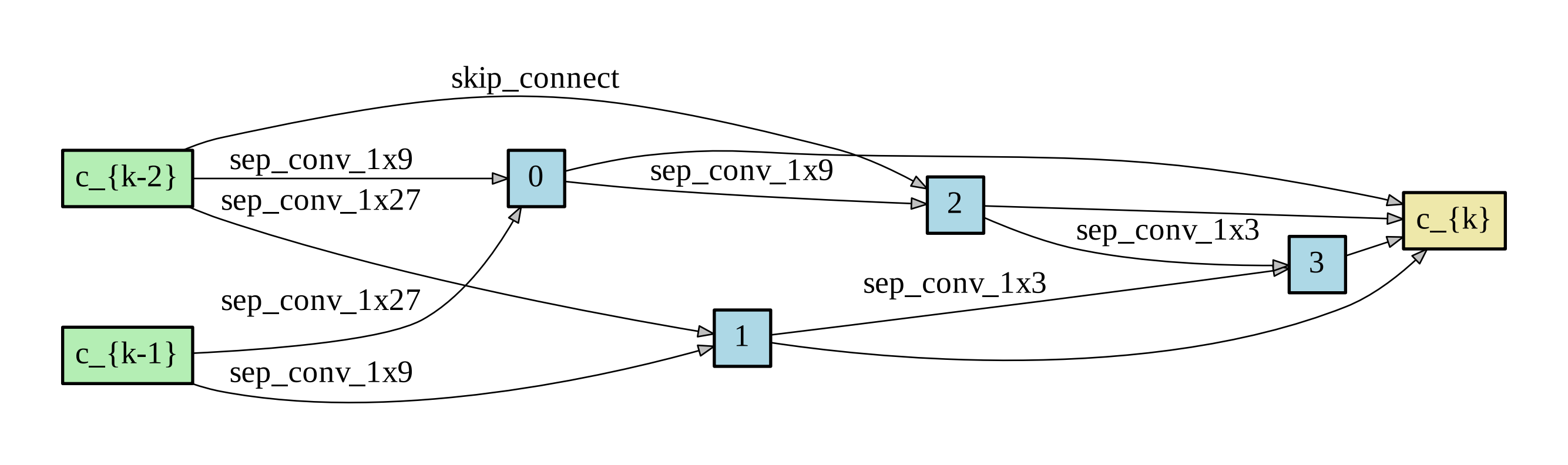}}
\subfigure[The reduction cell architecture of 2-lead data]{
\label{fig:2_lead_reduction} 
\includegraphics[width=3.0in]{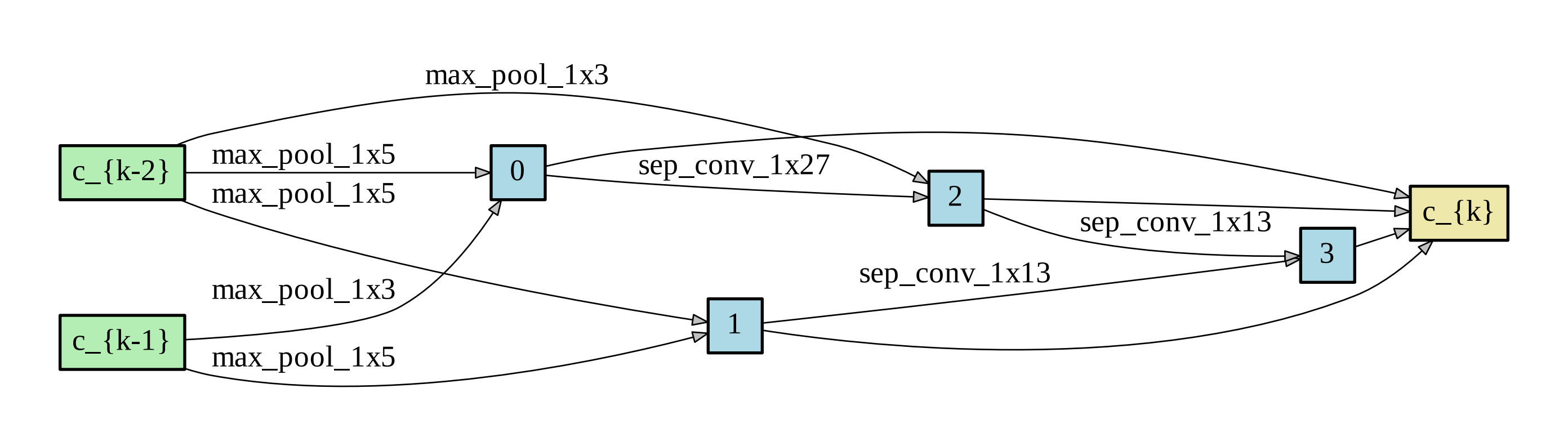}}
\caption{The cell architectures searched on 2-lead dataset}
\label{fig:2_lead_archi} 
\end{figure}

Results show that our Heart-Darts-base model achieves new state-of-the-art across all the metrics except precision. Specifically, the Heart-Darts-base model outperforms other models by approximately 0.4$\%$, 4$\%$, 2$\%$ in terms of accuracy, sensitivity, and f1-score, respectively,  for single-lead dataset. For the 2-lead dataset, our proposed model achieves approximately 0.3$\%$, 2$\%$, and 1$\%$ gain over other models in terms of accuracy, sensitivity, and f1-score, respectively. The promising outcomes of sensitivity and f1-score demonstrated that the Heart-Darts approach is more advantageous in data imbalance problems. Specifically, model architectures customized by auto-search methods may have stronger feature extraction capabilities than hand-crafted models for categories with less data. Additionally, the Heart-Darts-base model with shortcut connections yielded superior results, which demonstrates  the effectiveness of the proposed shortcut strategy.
It is worth mentioning that the 2-lead dataset further improves the performance of ECG classification compared with the single-lead as shown in Figure~\ref{fig:myplot}.


During the model evaluation phase, the detailed performance metrics of the Heart-Darts-base model for both datasets are presented in the confusion matrices, as shown in Tables~\ref{tab:Heart-Darts-1} and Tables~\ref{tab:Heart-Darts-2}.
For the single-lead dataset,  Heart-Darts achieved the global accuracy of 99.35$\%$ and the sensitivity of class F with the least data reach 85.71$\%$. For the 2-lead dataset, Heart-Darts achieved 99.43$\%$ accuracy and the sensitivity regarding the F class is 86.43$\%$. 
Figure~\ref{fig:2_lead_archi} shows the cell architecture of the Heart-Darts-base model and the whole process of cell architecture search costs 16 hours.

For model generalization validation, the results are presented in Table~\ref{tab:generalization}.
Note that the Heart-Darts-base model outperforms across all metrics except precision of the INCART database. Particularly, the proposed model outperforms the other two models by about 11$\%$ and 6$\%$ for the INCART database and by approximately 5$\%$ and 3$\%$ for the QT database on sensitivity and f1-score, respectively, which demonstrated that the Heart-Darts approach gains better generalization capability.

\section{Conclusion}
\label{sec:conclusion}
The purpose of this paper is to explore a promising CNN model being applied to ECG classification. To achieve this, we proposed a Heart-Darts approach by employing the Darts (i.e., an efficient neural architecture search method) to design an excellent CNN model. The Heart-Darts approach is the first application of Darts in ECG classification, which also proposes a novel model architecture to further improve the performance.
Specifically, Heart-Darts proposes a specific Darts for ECG analysis which automatically searches the CNN cell architecture.
After obtaining the cell architecture, Heart-Darts customizes a CNN model with a novel model architecture for ECG classification.  To validate the performance of the Heart-Darts-base model, we performed the model evaluation and generalization validation, both of them yielded appealing results.
Despite the promising
results we have obtained, the proposed Heart-Darts approach also suffers from several drawbacks. For example, the Darts search process relies heavily on the operation candidate set and the problem of model dimensionality constraints in Darts-based approaches.
These potential improvements are left for future work.

\section{Acknowledge}
This work is supported by the National Key Research and Development Project of China (No.2017YFB1002201), National Natural Science Fund for Distinguished Young Scholar (GrantNo. 61625204), the State Key Program of National Science Foundation of China (GrantNo. 61836006), and the Sichuan Science and Technology Program (NO. 2020YFG0323).





\printbibliography

@article{Goldberger2000PhysioBank,
  title={PhysioBank, PhysioToolkit, and PhysioNet: components of a new research resource for complex physiologic signals.},
  author={Goldberger, A. L. and Amaral, L. A. N. and Glass, L. and Hausdorff, J. M. and Ivanov, P. Ch. and Mark, R. G. and Mietus, J. E. and Moody, G. B. and Peng, C. K. and Stanley, H. E.},
  journal={Circulation},
  volume={101},
  number={23},
  pages={E215},
  year={2000},
}

@article{2020Heartbeat,
  title={Heartbeat classification using deep residual convolutional neural network from 2-lead electrocardiogram - ScienceDirect},
  author={ B, Zhi Li A  and  A, Dengshi Zhou  and  C, Li Wan  and  A, Jian Li  and  A, Wenfeng Mou },
  journal={Journal of Electrocardiology},
  volume={58},
  pages={105-112},
  year={2020},
}

@article{AAMI,
  title={Testing and reporting performance results of cardiac rhythm and ST segment measurement algorithms},
  author={Association for the Advancement of Medical Instrumentation (AAMI) (2008)},
pages={EC57},
  journal={American National Standards Institute, Inc. (ANSI), ANSI/AAMI/ISO  1998-(R)2008.},
  year={2008},
}

@article{Shanshan2017Heartbeat,
  title={Heartbeat classification using projected and dynamic features of ECG signal},
  author={Shanshan and Chen and Wei and Hua and Zhi and Li and Jian and Li and Xingjiao and Gao},
  journal={Biomedical Signal Processing \& Control},
  year={2017},
}

@article{2017Learning,
  title={Learning Transferable Architectures for Scalable Image Recognition},
  author={ Zoph, Barret  and  Vasudevan, Vijay  and  Shlens, Jonathon  and  Le, Quoc V },
  year={2017},
}

@article{Liu2018DARTS,
  title={DARTS: Differentiable Architecture Search},
  author={Liu, Hanxiao and Simonyan, Karen and Yang, Yiming},
  year={2018},
}

@article{zoph2016neural,
  title={Neural architecture search with reinforcement learning},
  author={Zoph, Barret and Le, Quoc V},
  journal={arXiv preprint arXiv:1611.01578},
  year={2016}
}

@inproceedings{zubair2016automated,
  title={An automated ECG beat classification system using convolutional neural networks},
  author={Zubair, Muhammad and Kim, Jinsul and Yoon, Changwoo},
  booktitle={2016 6th international conference on IT convergence and security (ICITCS)},
  pages={1--5},
  year={2016},
  organization={IEEE}
}

@article{acharya2017deep,
  title={A deep convolutional neural network model to classify heartbeats},
  author={Acharya, U Rajendra and Oh, Shu Lih and Hagiwara, Yuki and Tan, Jen Hong and Adam, Muhammad and Gertych, Arkadiusz and San Tan, Ru},
  journal={Computers in biology and medicine},
  volume={89},
  pages={389--396},
  year={2017},
  publisher={Elsevier}
}

@article{acharya2017automated,
  title={Automated detection of arrhythmias using different intervals of tachycardia ECG segments with convolutional neural network},
  author={Acharya, U Rajendra and Fujita, Hamido and Lih, Oh Shu and Hagiwara, Yuki and Tan, Jen Hong and Adam, Muhammad},
  journal={Information sciences},
  volume={405},
  pages={81--90},
  year={2017},
  publisher={Elsevier}
}

@inproceedings{liu2018progressive,
  title={Progressive neural architecture search},
  author={Liu, Chenxi and Zoph, Barret and Neumann, Maxim and Shlens, Jonathon and Hua, Wei and Li, Li-Jia and Fei-Fei, Li and Yuille, Alan and Huang, Jonathan and Murphy, Kevin},
  booktitle={Proceedings of the European Conference on Computer Vision (ECCV)},
  pages={19--34},
  year={2018}
}

@inproceedings{real2019regularized,
  title={Regularized evolution for image classifier architecture search},
  author={Real, Esteban and Aggarwal, Alok and Huang, Yanping and Le, Quoc V},
  booktitle={Proceedings of the aaai conference on artificial intelligence},
  volume={33},
  pages={4780--4789},
  year={2019}
}

@book{mendis2011global,
  title={Global atlas on cardiovascular disease prevention and control},
  author={Mendis, Shanthi and Puska, Pekka and Norrving, Bo and World Health Organization and others},
  year={2011},
  publisher={World Health Organization}
}

@book{kass2005basis,
  title={Basis and treatment of cardiac arrhythmias},
  author={Kass, Robert E and Clancy, Colleen E},
  volume={171},
  year={2005},
  publisher={Springer Science \& Business Media}
}

@article{sahoo2017multiresolution,
  title={Multiresolution wavelet transform based feature extraction and ECG classification to detect cardiac abnormalities},
  author={Sahoo, Santanu and Kanungo, Bhupen and Behera, Suresh and Sabut, Sukanta},
  journal={Measurement},
  volume={108},
  pages={55--66},
  year={2017},
  publisher={Elsevier}
}

@article{li2016ecg,
  title={ECG classification using wavelet packet entropy and random forests},
  author={Li, Taiyong and Zhou, Min},
  journal={Entropy},
  volume={18},
  number={8},
  pages={285},
  year={2016},
  publisher={Multidisciplinary Digital Publishing Institute}
}

@article{elhaj2016arrhythmia,
  title={Arrhythmia recognition and classification using combined linear and nonlinear features of ECG signals},
  author={Elhaj, Fatin A and Salim, Naomie and Harris, Arief R and Swee, Tan Tian and Ahmed, Taqwa},
  journal={Computer methods and programs in biomedicine},
  volume={127},
  pages={52--63},
  year={2016},
  publisher={Elsevier}
}

@article{li2016arrhythmia,
  title={Arrhythmia classification based on multi-domain feature extraction for an ECG recognition system},
  author={Li, Hongqiang and Yuan, Danyang and Wang, Youxi and Cui, Dianyin and Cao, Lu},
  journal={Sensors},
  volume={16},
  number={10},
  pages={1744},
  year={2016},
  publisher={Multidisciplinary Digital Publishing Institute}
}

@article{lee2017deep,
  title={Deep learning in medical imaging: general overview},
  author={Lee, June-Goo and Jun, Sanghoon and Cho, Young-Won and Lee, Hyunna and Kim, Guk Bae and Seo, Joon Beom and Kim, Namkug},
  journal={Korean journal of radiology},
  volume={18},
  number={4},
  pages={570--584},
  year={2017}
}

@article{krizhevsky2017imagenet,
  title={Imagenet classification with deep convolutional neural networks},
  author={Krizhevsky, Alex and Sutskever, Ilya and Hinton, Geoffrey E},
  journal={Communications of the ACM},
  volume={60},
  number={6},
  pages={84--90},
  year={2017},
  publisher={AcM New York, NY, USA}
}

@inproceedings{szegedy2015going,
  title={Going deeper with convolutions},
  author={Szegedy, Christian and Liu, Wei and Jia, Yangqing and Sermanet, Pierre and Reed, Scott and Anguelov, Dragomir and Erhan, Dumitru and Vanhoucke, Vincent and Rabinovich, Andrew},
  booktitle={Proceedings of the IEEE conference on computer vision and pattern recognition},
  pages={1--9},
  year={2015}
}

@inproceedings{he2016deep,
  title={Deep residual learning for image recognition},
  author={He, Kaiming and Zhang, Xiangyu and Ren, Shaoqing and Sun, Jian},
  booktitle={Proceedings of the IEEE conference on computer vision and pattern recognition},
  pages={770--778},
  year={2016}
}

@article{martis2013cardiac,
  title={Cardiac decision making using higher order spectra},
  author={Martis, Roshan Joy and Acharya, U Rajendra and Mandana, KM and Ray, Ajoy Kumar and Chakraborty, Chandan},
  journal={Biomedical Signal Processing and Control},
  volume={8},
  number={2},
  pages={193--203},
  year={2013},
  publisher={Elsevier}
}

@article{kiranyaz2015real,
  title={Real-time patient-specific ECG classification by 1-D convolutional neural networks},
  author={Kiranyaz, Serkan and Ince, Turker and Gabbouj, Moncef},
  journal={IEEE Transactions on Biomedical Engineering},
  volume={63},
  number={3},
  pages={664--675},
  year={2015},
  publisher={IEEE}
}

@article{yildirim2018novel,
  title={A novel wavelet sequence based on deep bidirectional LSTM network model for ECG signal classification},
  author={Yildirim, {\"O}zal},
  journal={Computers in biology and medicine},
  volume={96},
  pages={189--202},
  year={2018},
  publisher={Elsevier}
}

@article{yildirim2019new,
  title={A new approach for arrhythmia classification using deep coded features and LSTM networks},
  author={Yildirim, Ozal and Baloglu, Ulas Baran and Tan, Ru-San and Ciaccio, Edward J and Acharya, U Rajendra},
  journal={Computer methods and programs in biomedicine},
  volume={176},
  pages={121--133},
  year={2019},
  publisher={Elsevier}
}

@article{wang2020deep,
  title={A deep learning approach for atrial fibrillation signals classification based on convolutional and modified Elman neural network},
  author={Wang, Jibin},
  journal={Future Generation Computer Systems},
  volume={102},
  pages={670--679},
  year={2020},
  publisher={Elsevier}
}

@ARTICLE{Ye,
  author={Q. {Ye} and Y. {Sun} and J. {Zhang} and J. {Lv}},
  journal={IEEE Transactions on Parallel and Distributed Systems}, 
  title={A Distributed Framework for EA-Based NAS}, 
  year={2021},
  volume={32},
  number={7},
  pages={1753-1764},
  doi={10.1109/TPDS.2020.3046774}}

@inproceedings{lei-etal-2018-sequicity,
    title = "{S}equicity: Simplifying Task-oriented Dialogue Systems with Single Sequence-to-Sequence Architectures",
    author = "Lei, Wenqiang  and
      Jin, Xisen  and
      Kan, Min-Yen  and
      Ren, Zhaochun  and
      He, Xiangnan  and
      Yin, Dawei",
    booktitle = "Proceedings of the 56th Annual Meeting of the Association for Computational Linguistics (Volume 1: Long Papers)",
    month = jul,
    year = "2018",
    address = "Melbourne, Australia",
    publisher = "Association for Computational Linguistics",
    url = "https://www.aclweb.org/anthology/P18-1133",
    doi = "10.18653/v1/P18-1133",
    pages = "1437--1447",
}
\end{document}